\documentclass[11pt]{article}
\usepackage{amsthm}
\usepackage[top=2.7cm, left=2.9cm, right=3cm, bottom=2.7cm]{geometry}
\usepackage{amsmath}
\usepackage{times}
\usepackage{bm}
\usepackage{natbib}
\usepackage[utf8x]{inputenc}
\usepackage{amssymb}
\usepackage{bm}
\usepackage{float}
\usepackage[noabbrev,capitalise,nameinlink]{cleveref}
\usepackage{graphicx}
\usepackage{microtype}
\newcommand{\E}{{{E}}}
\newcommand{\cov}{{\mathrm{cov}}}

\newcommand{\var}{{\mathrm{var}}}

\newcommand{\I}{\mathbf{1}}

\usepackage{xcolor}

\graphicspath{{./art/}}

\usepackage[plain,noend]{algorithm2e}

\begin{document}

 \title{\bf Edge coherence in multiplex networks}
 \author{Swati Chandna 
\\
\textit{Department of Economics, Mathematics and Statistics}, \\
\textit{Birkbeck, University of London, UK}\\ s.chandna@bbk.ac.uk\\
\vspace{0.01mm}\\
Svante Janson
\\
\textit{Department of Mathematics, Uppsala University, 
Sweden}\\ svante.janson@math.uu.se\\
\vspace{0.01mm}\\
Sofia C. Olhede
\\
\textit{Institute of Mathematics}, \\ \textit{Ecole Polytechnique F\'ed\'erale de Lausanne,  Switzerland} \\ sofia.olhede@epfl.ch}
\date{}
 \maketitle
\begin{abstract}
{This paper introduces a nonparametric framework for the setting where multiple networks are observed on the same set of nodes, also known as multiplex networks. Our objective is to provide a simple parameterization which explicitly captures linear dependence between the different layers of networks. For non-Euclidean observations, such as shapes and graphs, the notion of ``linear'' must be defined appropriately. Taking inspiration from the representation of stochastic processes and the analogy of the multivariate spectral representation of a stochastic process with joint exchangeability of Bernoulli arrays, we introduce the notion of edge coherence as a measure of linear dependence in the graph limit space. Edge coherence is defined for pairs of edges from any two network layers and is the key novel parameter. We illustrate the utility of our approach by eliciting simple models such as a correlated stochastic blockmodel and a correlated inhomogeneous graph limit model.}
\end{abstract}

\section{Introduction}\label{sec:intro}
Graphs describe the interaction of objects. More recent work in statistics has acknowledged that graphs are fundamentally random objects and studied their random variation~\citep{kolaczyk2020averages}. If we can understand the variability of graphs, then it is natural to ask what different moments may be defined to characterise the dependence between graphs. For example~\citep{kolaczyk2020averages} has already addressed what the mean network of a set of networks corresponds to, which was also addressed by~\citep{lunagomez2020modeling} and others, for example~\citep{nielsen2018multiple,newman2018network,pamfil2020inference}. The question then naturally arises as to how one summarizes multiple networks when more than one network is under study. It is a question related to characterising multiplex and multilayer networks~\citep{pamfil2020inference, macdonald2020latent,bianconi2018multilayer}, or two graphs that are subject to graph matching~\citep{lyzinski2015spectral}. 

{There is already a growing popularity of multiplex and dynamic network data structures \cite[e.g.][to name a few]{lei2020consistent, lunagomez2020modeling, zhang2017finding} where two or more networks are observed. A variety of parametric probabilistic models and estimation methods have been proposed in the literature, e.g. latent space approaches such as \cite{arroyo2021inference}, \cite{Gollini16} and more recently \cite{macdonald2020latent}. While it is extremely important to develop models which can sufficiently capture the complex structure observed in such data, it is equally important to quantify dependence between pairs of networks and have a framework which allows for the empirical description of  correlated networks.}

The closest areas of statistics to take inspiration from when studying dependence measures for graphs are dependence measures for other non-Euclidean objects such as in statistical shape analysis~\citep{dryden2016statistical}, summary statistics of random matrices~\citep{ramsay1984matrix} as well as representations of stochastic processes\\
~\citep{adler2009random}. Most stochastic processes admit a representation in terms of a suitable representation theorem, such as the spectral representation theorem~\citep{adler2009random}, or the chaotic representation property~\citep{nualart2000chaotic}. The spectral representation of stochastic processes will inspire our choice of joint representation of multiple networks. In the case of networks we have to make some decisions about a suitable measure of correlation--and determine if we want to make connections with linear filtering of stochastic processes (a superficial similarity that will in general break down as linear operations on adjacency matrices do not produce new adjacency matrices of simple graphs),  or whether we want a measure that relates to conditional probability of observing an edge in one graph, given we did so in another graph. We resolve these issues by studying the multivariate Bernoulli distribution with the graph limit representation of networks, \citep{lovasz2012large}.

{The key to specifying dependence measures is their ability to characterise the intrinsic structure of the objects they describe.} 
{We discuss the notion of dependence from a perspective of defining appropriate fundamental representations, which, we believe is crucial to the study of multiple networks.}
To specify dependence between two (or more) binary networks, we study the bivariate (or multivariate) Bernoulli distribution~\citep{teugels1990some} with a view to understand how this, in general, is linked to the notion of graph limits~\citep{lovasz2012large} { obtained under the framework of joint exchangeability. For the sake of interpretation, joint exchangeability may be considered analogous to joint
locally stationary stochastic processes~\citep{tong1973some}, that require a
representation in {\em the same} locally oscillatory family.}
This naturally yields a dependence measure that we refer to as {\em edge coherence} arising as a notion of correlation directly from the graph limit representation of a network.
{For a pair of networks on the same set of $n$ nodes, edge coherence is an $n \times n$ matrix with the $ij$th entry 
quantifying linear dependence between edges in the corresponding adjacency matrices.}

{ Our intuition for multivariate analysis is based on the Gaussian case, which in turn relies on Isserlis'~\citep{isserlis1918formula} formula. This no longer holds for Bernoulli random variables and has a number of consequences. For example, to ensure that $d$ Bernoullis are independent, one has to prescribe that the coefficients of the interaction functions are all zero; unlike in the Gaussian case, this is a hierarchy of interaction coefficients~\citep[Theorem 3.1]{dai2013multivariate} involving cross-product ratios, that are directly interpretable in terms of log-odds ratios~\citep{whittaker1990graphical}.  
 Further, the validity of a multivariate Bernoulli model is much harder to determine
~\citep{lovison2006matrix,teugels1990some,huber2017bernoulli}, and cannot simply be built up from a bivariate understanding.}
Our derivations show that the assumption of a common latent vector underlying all graphs, is key to ensuring a simple and valid joint representation, as is the case in the study of non-stationary processes.
Further, we establish links with decorated graphs~\citep{lovasz2010limits} and directed graphs~\citep{janson2008graph}, as such data also corresponds to two Bernoulli random variables per prospective edge.

To illustrate the properties and utility of the proposed framework, we discuss how some simple network models generate dependence. Specifically, we describe a correlated stochastic blockmodel and more generally, a correlated graph limit model with suitable parameterizations, and note how the commonly used choice of scaling for sparse networks, has important implications. Our proposed approach not only builds on classical multivariate methods for network analysis but also aids in interpretation of linear dependence between graphs under the widely used framework of joint exchangeability.

\section{Notation and the Bivariate Bernoulli Distributions}\label{sec:notation}
We shall start by assuming that we observe two graphs ${\cal G}_1$ and ${\cal G}_2$ with the same set of nodes denoted as $\{1,\dots, n\}$ {or $[n]$}. 
{Let $\{A_{ij}^{(1)}\}$ and $\{A_{ij}^{(2)}\}$ denote the two sets of edge variables representing two symmetric arrays and let $\underline{A}_{ij}=\begin{pmatrix} A_{ij}^{(1)} & A_{ij}^{(2)}\end{pmatrix},$ where $(i,j)\in[n]\times [n]$ denote the corresponding two-vector valued observation. In general, we may simultaneously consider $d$ networks and likewise define the corresponding $d$--variate vector for each $(i,j)$ as $\underline{A}_{ij}=\begin{pmatrix} A_{ij}^{(1)} & A_{ij}^{(2)} & \dots & A_{ij}^{(d)}\end{pmatrix}$.}

{From the characterization of any bivariate Bernoulli vector studied by~\citep{teugels1990some}, we know that three parameters, specifically, two (marginal) success probabilities of observing an edge in either graph that is $\E\{A_{ij}^{(1)} \}$ and $\E\{A_{ij}^{(2)} \}$, in combination with a co-dependence measure (which may or may not be centered), are required for full specification.}
If we observe $d$-variate Bernoullis, then Teugels observes that $2^d-1$ (centered or non-centred) parameters must be specified to provide the distribution (a parameter per probability of the possible outcomes, with one degree of freedom removed since all distributions are normalized to area $1$). To specify co-dependence between the two random variables that are marginally Bernoullis we need to understand an appropriately chosen moment.
A simple unnormalized co-dependence measure is obtained by studying the element-wise (Hadamard) matrix product of $A^{(1)}$ and $A^{(2)}$ given by
\begin{equation*} \left(A^{(1)}\circ A^{(2)}\right)_{ij}=A_{ij}^{(1)}A_{ij}^{(2)}, (i,j)\in [n] \times [n].\end{equation*}
It is clear that the expectation of the Hadamard product corresponds to a joint probability, i.e.
\begin{equation}
\label{eqn:hadamard}
\E\left\{ \left(A^{(1)}\circ A^{(2)}\right)_{ij}\right\}=\Pr \left\{ A_{ij}^{(1)}=1\quad {\mathrm{and}} \quad A_{ij}^{(2)}=1 \right\}. \end{equation}
This trivial observation yields two (statistical) interpretations of the Hadamard product of adjacency matrices; either as a joint probability of two events occurring (which could easily be manipulated to conditional probabilities, say what is the probability ${\cal G}_2$ has an edge between $i$ and $j$ given 
${\cal G}_1$ does) or as an uncentred second moment, naturally transformed into a correlation of edge variables by  centering and normalising.

A correlation requires the subtraction of the product of first moments, but both the expectation of the centred and uncentred product naturally satisfy the Cauchy--Schwarz inequality, this leading in each case to a natural choice of normalization. To progress beyond recognizing the moments of~\eqref{eqn:hadamard} we need to introduce a model, and here we shall progress in the framework of scaled graph limits~\citep{bickel2009nonparametric,bollobas365metrics}.  

\section{Scaled Graph Limits}\label{sec:scaledgraphs}
\subsection{Joint exchangeability}
A key early assumption to allow for the study of sparse graphs 
is to posit a model of a scaled graph limit for the observations. To start from our understanding of dense graphs (a graph where the probability of observing a randomly selected edge remains order one even as we observe larger graphs), let us recall the definition of an exchangeable array~\citep{lauritzen2008exchangeable}, namely that an (infinite) array $A$ is jointly exchangeable if for any permutation $\pi$ it satisfies $(A_{\pi(i) \pi(j)})\overset{d}{=} (A_{ij}),$
where $\overset{d}{=}$ denotes equivalence in distribution. This definition is normally posed for infinite arrays, as otherwise we need to use the notion of finite exchangeability, rather like a stationary stochastic process is related to a cyclostationary stochastic process (this describing the behaviour of the ``boundary'' of any array).
A consequence of joint exchangeability 
is the 
Aldous--Hoover representation theorem~\citep[Thm 7.22]{kallenberg2006probabilistic}.
It states that if $(A_{ij})$ is {{jointly}} exchangeable then there exists a graph limit 
$f(x,y;\alpha)$, and independent latent variables $\alpha$ and $\{\xi_i\}$, so that
\begin{equation}
    \label{eqn:AHthm}
    A_{ij}\,|\, \alpha, \xi \overset{{\mathrm{ind}}}{\sim }
    {\mathrm{Bernoulli}}\{ f(\xi_i,\xi_j;\alpha)\},\quad i<j,
\end{equation}
and the array is completed symmetrically setting 
$A_{ii}=0$ to correspond to a simple graph. The $\{A_{ij}\}$ are conditionally independent given $\xi$ and $\alpha$, as we indicate using ${\mathrm{ind}}$.
{{The assumption of joint exchangeability for a vector-valued binary array which we denote as $\mathbf{A}$ likewise implies distributional invariance under permutation of labels i.e. for any permutation $\pi$ it follows
that}}
\begin{equation}
\label{eqn:perminv}
(\underline{A}_{\pi(i) \pi(j)})\overset{d}{=} (\underline{A}_{ij}).\end{equation}

The Aldous--Hoover theorem~\citep[Thm 7.22]{kallenberg2006probabilistic}
applies in this setting too, and shows that 
for an infinite array of $d$-vectors satisfying \eqref{eqn:perminv},
there exist  
independent latent variables $\alpha$ and $\{\xi_i\}$, so that
(generalizing \eqref{eqn:AHthm})
 the vectors $\underline{A}_{ij}$
are conditionally independent with $d$-variate Bernoulli distributions depending on
$\alpha$, $\xi_i$ and $\xi_j$, i.e.
\begin{equation}
    \label{eqn:AHthmvec}
    \underline{A}_{ij}\,|\, \alpha, \xi \overset{{\mathrm{ind}}}{\sim }
    {\mathrm{MultBern}}_d\bigl\{\underline{f}(\xi_i,\xi_j;\alpha);{f}^{(12)}(\xi_i,\xi_j;\alpha),\dots, {f}^{(d-1\,d)}(\xi_i,\xi_j;\alpha),\dots \bigr\},
\end{equation}
where ${\mathrm{MultBern}}_d\{ \}$ is the $d$-variate Bernoulli
distribution~\citep{teugels1990some}, here specified by 
the $2^d-1$ cross-moments
${f}^{(m_1,\dots, m_l)}$, which are enumerated by the set of non-empty
subsets $\{m_1,\dots, m_l\}$ of $\{1,\dots, d\}$, and we write
\begin{equation}
\underline{f}(x,y;\alpha)=\begin{pmatrix} {f}^{(1)}(x,y;\alpha) & \dots &
{f}^{(d)}(x,y;\alpha)\end{pmatrix}^T,
\end{equation}
for the vector of the $d$ means. To construct a model for sparse graphs for a finite $n$-node sample, following~\citep{bickel2009nonparametric,bollobas365metrics}, we modify~\eqref{eqn:AHthm} by introducing a scale factor $\rho_n>0$, and let the distribution
be specified by
\begin{equation}
    \label{eqn:AHthmsp}
    A_{ij}\,|\, \alpha, \xi \overset{{\mathrm{ind}}}{\sim }
    {\mathrm{Bernoulli}}\{ \rho_n f(\xi_i,\xi_j;\alpha)\},\quad 1\leq i<j\leq n.
\end{equation}
Here it is assumed that $f(x,y)$ is a non-negative symmetric function.
It is often convenient to further assume that $\|f\|_1=1$, since
it ensures that $\rho_n$ is a marginal edge probability
and makes $\rho_n$ identifiable from the observed graphs. 
If
$\rho_n$ is shrinking with $n$, then if $f(x,y)$ is bounded, eventually
$\rho_n \|f\|_{\infty}$ will be less than unity; various other fixes can be
discussed \citep[cf. e.g.][(2.3) and Remark 2.4]{sj178}.
{{Given this,}} we will simply assume that $f(x,y)$ is bounded
and that $\rho_n \|f\|_{\infty}\le 1$. In what follows, we shall suppress the indexing on the random variable $\alpha$. This corresponds 
to assuming that we observe dissociated random variables.

\subsection{Multiplex graphs with $d\geq 2$ layers}
To elicit a model for $d$ layers of sparse graphs on a finite number of $n$ nodes, we take 
inspiration from~\eqref{eqn:AHthmvec} and \eqref{eqn:AHthmsp} 
and assume that
for each $(i,j)$ with $1\leq i<j\leq n$, a vector $\underline{A}_{ij}$ 
is generated such that
\begin{equation}
    \label{eqn:AHthmspvec}
    \underline{A}_{ij}\,|\, \xi \overset{{\mathrm{ind}}}{\sim }
    {\mathrm{MultBern}}_d\{ \rho_n\underline{f}(\xi_i,\xi_j);\rho_n^2{f}^{(12)}(\xi_i,\xi_j),\dots, \rho_n^2{f}^{(d-1\,d)}(\xi_i,\xi_j),\dots \},
\end{equation}
where the $d$-variate Bernoulli distribution ${\mathrm{MultBern}}_d\{ \}$ 
 is specified by $2^d-1$ parameters, which we separate into 
a mean vector $\rho_n \underline{f}(\xi_i,\xi_j)$
and a higher order structure. 
The choice of scaling of the second order structure, which in general may be denoted as $\rho^{\gamma}_{n}$ for $\gamma\in {\mathbb{Q}}_{\geq 0}$ is defined for networks with an increasing number of nodes $n$. By assuming a square power of $\rho_n$ as $n$ increases i.e. with $\gamma=2$ we expect covariant behavior of the covariance i.e. the joint probability of an edge in (any) two networks scales as the product of edge probabilities in each network. This is natural for Gaussian data but can be problematic for sparse networks. Alternatively, $\gamma$ may be chosen to be one to allow concurrent edges at the same rate as in the individual networks.
Note that if the moments ${f}^{(m_1,\dots, m_l)}$ are compatible, then
so are the scaled moments $\rho_n^l{f}^{(m_1,\dots, m_l)}$,
since we may obtain the sparse version by first generating a dense array
satisfying \eqref{eqn:AHthmvec}
and then obtain the distribution in~\eqref{eqn:AHthmspvec} by thinning,
i.e., (in this case) 
multiply all $A_{ij}^{(m)}$ by independent ${\mathrm{Bern}}(\rho_n)$ random
variables. 
{Such thinning is independent across layers; we could also use a single thinning operation across all layers. This does not matter marginally and will only make a difference in terms of co-dependence.}
Also, note that the latent variables are chosen to be common across the $d$ graph layers. Let  $p^{(m)}_{ij}=p^{(m)}(\xi_i,\xi_j)=\rho_n f^{(m)}(\xi_i,\xi_j),$
 denote the edge probability for node pair $(i,j)$ in the $m$th graph, where $m \in \{1,\ldots,d\}$.

The first thing to note in this setting is that for Bernoulli random variables, the
distribution is not specified by the $d$ first order moments
and $\binom{d}{2}$ covariances, as it is for Gaussians where then any
cross-moment can be computed using Isserlis'
theorem~\citep{isserlis1918formula}. Care of course
needs to be taken not to under-specify (or over-specify) a given
distribution. For Bernoulli random variables,
in addition to the means and covariances, an additional $2^d-1-d-\binom{d}{2}$ 
moments need to be specified.
Of course, when $d=2$, the total number of parameters $2^d-1=3=d+\binom{d}{2}$ 
 and thus, no additional parameters are required. 
The second key observation is that \eqref{eqn:AHthmspvec} specifies a homogeneous scaling across edge probabilities in the $d$ graph layers. Scaled graphons were introduced in~\citep{bollobas365metrics} so that the behaviour of finite sized networks could be studied. For very sparse graphs, much of the structural behaviour is determined by their degree of sparsity. We chose to use a single scaling {across all graph layers} to simplify the modelling, but fully acknowledge that this is a constraint.

Now, from~\eqref{eqn:AHthmspvec} we note that the marginal correlation of
$A_{ij}^{(m_1)}$ with $A_{kl}^{(m_2)}$ is specified 
through forward
modelling and using the theorem of total covariance
it follows that
\begin{align*}
    \cov\{ A_{ij}^{(m_1)},A_{kl}^{(m_2)}\}
    &= \cov_{{\xi}}\{ \E_{A|{\xi}}\{A_{ij}^{(m_1)}\},\E_{A|{\xi}}\{A_{kl}^{(m_2)}\}\}
+ \E_{{\xi}}\{\cov_{A|{\xi}}\{A_{ij}^{(m_1)},A_{kl}^{(m_2)} \}\},
\end{align*}
where $1\leq i<j\leq n$ and $1\leq k<l\leq n$, as well as $1\leq m_2,\,m_2\leq d$.
We see that this decomposition directly allows us to characterise two types
of dependence; dependence that is intrinsically bivariate specified from the
latent variables $\xi$, 
but also dependence specified conditionally on the latent variables. If
we observe only one array then one can use this decomposition to show that
if the array is dissociated then $A_{ij}$ and $A_{il}$ have non-negative
covariance bounded above by $(1/2) \var\{A_{ij}\}$; this limits also the
correlation of this variable. If the array is associated, then the
covariance of $A_{ij}$ and $A_{il}$ is strictly
positive~\citep{aldous,bassan1998finite}. 
Furthermore, for two arrays,
requiring that $\cov_{A|\xi}\{A_{ij}^{(1)},A_{ij}^{(2)}\}=0$ is equivalent
to requiring $f^{(12)}(x,y)=f^{(1)}(x,y)f^{(2)}(x,y)$.
This may
be realistic in some scenarios, but as an assumption pushes all correlation
between processes into the latent process. This may constrain what marginal
probability functions we can use, not to mention what correlations are
achievable. 

Why does this matter? The problem of covariance is that while forward modelling of the dependence of $A_{ij}$ is trivial, inverse specification of an arbitrary correlation for marginally Bernoulli variables is not~\citep{chaganty2006range}, and at times we will wish to achieve a negative correlation between edges, especially if edges require the same resource to form. 
For the purposes of estimation with only one realization of $
\underline{A}_{ij}$ for $1\leq i<j\leq n$ where $n\in{\mathbb{N}}$ we can only
implement conditional inference, conditionally on the realized value of
$\alpha$, no matter if $n\gg 1$. This is consistent with recent results in
estimating subgraph densities from noisy
networks~\citep{chang2020estimation}. 

We note that the model elicited above can
alternatively be specified in terms of centred 
moments, or directly by the vector of the $2^d$ (conditional) probabilities as 
\begin{equation} \label{sj1}
\Pr\{\underline{A}_{ij}=\underline{a}\mid \xi\}
=p_{\underline{a}}(\xi_i,\xi_j),
\qquad
\underline{a}=(a_1,\dots,a_d)\in\{0,1\}^d,
\end{equation}
{where $\underline{a}$ denotes the presence or absence of an edge between nodes $i$ and $j$ in graphs $\{1,\ldots,d\}$,}
see~\citep{teugels1990some}, where conversion
formulas are given. 
The functions 
${f}^{(m_1,\dots, m_l)}(x,y)$ are symmetric in $(x,y)$ i.e.
${f}^{(m_1,\dots, m_l)}(x,y)={f}^{(m_1,\dots, m_l)}(y,x)$,
and similarly for $p_{\underline{a}}$ in \eqref{sj1}.
Moreover,
the moments ${f}^{(m_1,\dots, m_l)}$
must be consistently specified
\citep[e.g.]{huber2017bernoulli,chaganty2006range}.
Other discussions of the multivariate Bernoulli distribution can be found
in~\citep{joe1997multivariate,whittaker1990graphical}. 

For simplicity, we turn our focus to multiplex networks
in the simplest setting of $d=2$ layers.
In this case we have  for $1\leq i<j\leq n,$
\begin{equation}
    \label{eqn:AHthmspvec2d}
    \underline{A}_{ij}\,|\, \xi_{i},\xi_{j} \overset{{\mathrm{ind}}}{\sim }
    {\mathrm{MultBern}}_2\{ \rho_n \underline{f}(\xi_i,\xi_j);\rho_n^2{f}^{(12)}(\xi_i,\xi_j) \},
\end{equation}
where ${\mathrm{MultBern}}_2\{ \}$ is denoting the two-vector distribution, and $\underline{f}(x,y)=\begin{pmatrix} {f}^{(1)}(x,y) &
{f}^{(2)}(x,y)\end{pmatrix}^T.$
This distribution for a fixed $i$
and $j$ is specified by three moments, chosen as the two means, and the cross-moment as a
co-dependence measure. In some settings log odds ratio are advised for this
modelling~\citep{whittaker1990graphical}. 
Alternatively, we may specify the distribution by four functions $p_{\underline{a}}$
as in \eqref{sj1}; these sum to 1, so we only have to specify three of them.
There are simple algebraic relations between these functions and the 
functions in \eqref{eqn:AHthmspvec2d} given by
\begin{align}
\rho_nf^{(1)}(x,y)&=p_{10}(x,y)+p_{11}(x,y)\label{pf1}\\
\rho_nf^{(2)}(x,y)&=p_{01}(x,y)+p_{11}(x,y)\label{pf2}\\
\rho_n^2{f}^{(12)}(x,y)&=p_{11}(x,y) 
\label{pf12}
.\end{align}

This representation allows us to further note linkages between~\eqref{eqn:AHthmspvec} 
and the representation of directed graphs~\citep{janson2008graph}. 
Here directed graphs are parameterized via three independent measurable functions, that we shall
denote $g_{00}(x,y;\alpha),$ $g_{01}(x,y;\alpha)$ and $g_{10}(x,y;\alpha)$ where a fourth 
function is defined by the theorem of total probability implicitly as
$g_{11}(x,y;\alpha)=1-g_{00}(x,y;\alpha)-g_{01}(x,y;\alpha)-g_{10}(x,y;\alpha)$. 
These functions correspond to the functions $p_{kl}(x,y;\alpha)$
 above; 
however, the symmetry requirements on the 
functions $g_{kl}(x,y;\alpha)$ are different than those imposed on $p_{kl}(x,y;\alpha)$,
because directed edges correspond to a kind of antisymmetric 
relations between nodes
rather than symmetric relations,
(e.g. see
\cite{janson2008graph} for details).

\section{Edge Coherence}\label{sec:netcoh}
We should start by noting that the terminology ``coherence'' has 
been used in another context for networks~\citep{ding2015network,patterson2011network} and has a different meaning. Our motivation for 
re-using the term comes from stochastic processes, especially time series. Some care must be used in understanding the analogy but not over-interpreting it. For a time series, coherence captures linear dependence between zero-mean stochastic processes, and convolutions (``filtering'') in the temporal domain, correspond to linear relationships in the frequency domain. Of course we cannot entertain linear relationships between networks--as linear combinations of adjacency matrices of simple graphs do not yield adjacency matrices of simple graphs. Linear combinations of graph limits with positive weights do yield new graph limit functions, as long as overall scaling is treated carefully and the graph limit does not when scaled exceed unity. 
To write down an appropriate definition of a dependence measure we need to determine what we are quantifying. The marginal correlation between edges in two networks is given by
\begin{equation}
    \sigma_{ij}=\frac{\E\{A_{ij}^{(1)}A_{ij}^{(2)}\}-\E\{A_{ij}^{(1)}\}\E\{A_{ij}^{(2)}\}}{[{\var\{A_{ij}^{(1)}\}\var\{A_{ij}^{(2)}}\}]^{1/2}}.
\end{equation}
Of course estimating this quantity will require some additional assumptions,
and we need to specify what expectations we are calculating. Calculating the
expectations conditionally on $\xi$, we get 
the conditional correlation
\begin{align}\label{eqn:coh}
r(\xi_{i},\xi_{j})&=\frac{\E_{A|{\xi}}\{A_{ij}^{(1)}A_{ij}^{(2)}\}-\E_{A|{\xi}}\{A_{ij}^{(1)}\}\E_{A|{\xi}}\{A_{ij}^{(2)}\}}{[{\var_{A|{\xi}}\{A_{ij}^{(1)}\}\var_{A|{\xi}}\{A_{ij}^{(2)}\}}]^{1/2}} \nonumber \\
    &=\frac{\rho_n\left[{f}^{(12)}(\xi_i,\xi_j)-f^{(1)}(\xi_i,\xi_j) f^{(2)}(\xi_i,\xi_j)\right]}{[{f^{(1)}(\xi_i,\xi_j)\left\{1-\rho_nf^{(1)}(\xi_i,\xi_j)\right\}f^{(2)}(\xi_i,\xi_j)\left\{1-\rho_nf^{(2)}(\xi_i,\xi_j)\right\}}]^{1/2}}.
\end{align}
Note that, in eliciting the multiplex model with two layers in \eqref{eqn:AHthmspvec}, we had chosen the cross-moment $\rho_n^2f^{(12)}$ as the co-dependence
measure. Equivalently, we may use the coherence function as defined above.

We now ask, what does it mean if 
$r(\xi_i,\xi_j)=0$ or if $|r(\xi_i,\xi_j)|=1$? 
In the former case the conditional (on $\xi$) probability of an edge in both
the layers simply corresponds to the product of the marginal probabilities,
as would be expected.
On the other hand, perfect correlation 
between Bernoulli variables means that they coincide. 
This means that the marginal probabilities are the same, 
so that $f^{(1)}(\xi_i,\xi_j)=f^{(2)}(\xi_i,\xi_j)=f(\xi_i,\xi_j)$, 
and, moreover, that
the expected value of the Hadamard product is the same as the expected value
of either of the two individual adjacencies, 
i.e.,
$\rho_n^2 f^{(12)}(\xi_i,\xi_j)=\rho_nf(\xi_i,\xi_j)$,
or
\[\rho_n f^{(12)}(\xi_i,\xi_j)=f(\xi_i,\xi_j).\]
Similarly, $r(\xi_i,\xi_j)=-1$ is possible only if
$A_{ij}^{(1)}=1-A_{ij}^{(2)}$ a.s.,
and thus $\rho_n\bigl(f^{(1)}(\xi_i,\xi_j)+f^{(2)}(\xi_i,\xi_j)\bigr)=1$,
which is impossible in the sparse case (for large $n$ and thus small $\rho_n$).

In general,
for a bivariate Bernoulli vector to be well-defined, the moments characterizing it must satisfy certain inequalities. Specifically, we can only specify $(\rho_{n}, \underline{f}(x,y), r(x,y) )$ such that \citep{chaganty2006range} 

{\small \begin{multline}
\max_{} \left\{-\biggl({\frac{\rho_n^2{f}^{(1)}(x,y){f}^{(2)}(x,y)}
{(1-\rho_n{f}^{(1)}(x,y))(1-\rho_n{f}^{(2)}(x,y)) }\biggr)^{1/2}  },
     -\biggl({\frac{(1-\rho_n{f}^{(1)}(x,y))(1-\rho_n{f}^{(2)}(x,y
     ))}{\rho_n^2{f}^{(1)}(x,y){f}^{(2)}(x,y)
          }}\biggr)^{1/2}\right\} \\ 
    \leq r(x,y)\leq  \\
     \min_{} \left\{\biggl({\frac{{f}^{(1)}(x,y)(1-\rho_n{f}^{(2)}(x,y))}{ {f}^{(2)}(x,y)(1-\rho_n{f}^{(1)}(x,y)) }\biggr)^{1/2}}, \biggl({\frac{(1-\rho_n {f}^{(1)}(x,y)){f}^{(2)}(x,y)}{{f}^{(1)}(x,y)(1-\rho_n {f}^{(2)}(x,y) )}\biggr)^{1/2}}\right\},\; \forall \; (x,y)\in [0,1]^2.
     \label{corrlimits}
\end{multline}}
Therefore, not all possible choices can be realized
in~\eqref{eqn:AHthmspvec2d}. Negative correlation can be problematic, as observed in other settings, see for example~\citet{mckenzie2003ch} or \citet{andel2010bernoulli}.  

For a sparse graph, $r(\xi_{i},\xi_{j})$ is not an order one quantity, which will continue to decrease with $n$, that should eventually be estimated as zero. We therefore define the edge coherence function to be given by
\begin{equation}
\label{eqn:coherence}
r_{0}(\xi_i,\xi_j)=r(\xi_i,\xi_j)/\rho_n,\end{equation}
as a proper object behaving consistently with increasing $n$ as long the homogeneous scaling assumption in $\rho_n$ is correct.
Alternatively, if cross-moments in \eqref{eqn:AHthmspvec}
       were chosen to scale as $\rho$ i.e. $\gamma=1$, 
then $r(\xi_i,\xi_j)$ is an $O(1)$ quantity, and we 
define edge coherence by 
$r_0(\xi_i,\xi_j)=r(\xi_i,\xi_j)$
instead of  \eqref{eqn:coherence}.

Note that our choice of coherence in~\eqref{eqn:coherence} is centred, i.e., we chose to remove a product of the means of both arrays $\{A_{ij}^{(1)}\}$ and $\{A_{ij}^{(2)}\}$, and study fluctuations around this product of means, suitably standardized.
If $\rho_n\rightarrow 0$ then
\begin{align}
\label{eqn-sparsecorr}
  {r_{0}(\xi_i,\xi_j)}=\frac{{f}^{(12)}(\xi_i,\xi_j)-f^{(1)}(\xi_i,\xi_j) f^{(2)}(\xi_i,\xi_j)}{\{{f^{(1)}(\xi_i,\xi_j)f^{(2)}(\xi_i,\xi_j)}\}^{1/2}}\{1+o(1)\},
\end{align}
and apart from the mean correction,
this form resembles the coherence for a stochastic process~\citep{adler2009random}. 
There are other ratios we could have defined in~\eqref{eqn-sparsecorr}. Notice that $\E_{A|{\xi}}\{A_{ij}^{(1)}A_{ij}^{(2)}\}=\Pr\{A_{ij}^{(1)}=1\,\cap\, A_{ij}^{(2)}=1\,|\,\xi \},$
and so when $\rho_n$ is appreciable, and the marginal probabilities are the
same, then~\eqref{eqn-sparsecorr} 
{ before the mean was subtracted} can be interpreted as a mean (and sparsity) corrected {partial correlation between edges}. { For stochastic processes, the interpretation of coherence is in terms of linear prediction of zero-mean random variables, which is not the case for networks, where the mean structure is important. }

Finally, we might ask, how do we deal with the notion of coherence for more than two graphs? We can collect any of the $\binom{d}{2}$ pairwise measures of dependence and collect them in a data structure. This will not completely characterise the multivariate structure of the $d$ variables, but will provide a simple summary {of their dependence structure}, like using correlation for non-Gaussian but Euclidean data structures.

\section{Models That Generate Correlated Graphs}\label{sec:corr-net}
\subsection{Correlated Stochastic Blockmodel}\label{subsec:corrstochblock}
{We elicit the correlated stochastic blockmodel (CSBM) for the simplest case of $d=2$ layers. The general case follows by applying the definitions here to all possible pairs of graphs. 
Let $K$ denote the number of blocks in each graph and let $z_i\in\{1,\dots,K\}$ where $|\{i,\;z_i=a\}|=h_a$, provide the block label to which node $i$ belongs. Normally $\{z_i\}$ are unobserved and part of the inference problem is determining how to cluster the observed network (e.g. \cite{olhede2014network}). Let  $\Theta^{(1)}=(\theta^{(1)}_{ab})$ and likewise $\Theta^{(2)}=(\theta^{(2)}_{ab})$ where ${(a,b)\in [K]\times [K]}$, denote the $K \times K$ edge-probability matrix corresponding to $A^{(1)}$ and $A^{(2)}$, respectively, i.e. 
\begin{equation}
P(A^{(l)}_{ij}=1|z_{i}=a,z_{j}=b)=\theta^{(l)}_{ab} , \hspace{3mm} l\in \{1,2\}.
\end{equation}
Given nodes $i$ and $j$ such that $z_{i}=a$ and $z_{j}=b$, let $r_{ab}$ denote the coherence between blocks $a$ and $b$, i.e. $r_{ab}=\{r_{ij}|z_{i}=a, z_{j}=b\},$
where $r_{ij} \equiv r(\xi_{i},\xi_{j})$ } given by \eqref{eqn:coh} is the conditional correlation between $A^{(1)}_{ij}$ and $A^{(2)}_{ij}$. Then $R=(r_{ab}),$ denotes the full $K \times K$ coherence matrix specifying conditional correlation between the two network layers. Thus, given $K$, the correlated stochastic blockmodel is parameterized via  $(z,\Theta^{(1)}, \Theta^{(2)}, R)$ where following the time series terminology, we refer to $\Theta^{(l)}, l=1,2$ as the two auto-interaction parameters and $R$ as the coherence parameter. 
Alternatively, we may parameterize the model via the uncentred joint moment
$\varrho_{ab}=\E\{A^{(1)}_{ij} A^{(2)}_{ij} \}$, \citep[e.g. as in][]{pamfil2020inference}. This is naturally linked to the coherence parameter since $r_{ab}=({\varrho_{ab}-\theta^{(1)}_{ab}\theta^{(2)}_{ab}})/({\theta^{(1)}_{ab}\{1-\theta^{(1)}_{ab}\}}{\theta^{(2)}_{ab}\{1-\theta^{(2)}_{ab}\}})^{1/2}$. 

Given the model parameters as described above, a pair of correlated SBM networks may be generated easily  following ~\citep{lunn1998note}.  For simplicity, assume that $\Theta^{(1)}=\Theta^{(2)}=(\theta_{ab})$. Then $(A^{(1)},A^{(2)})$ are networks from the corresponding bivariate SBM, if for $i< j$ independently, we set
\begin{align}\label{eqn:gen}
A_{ij}^{(1)}&=\{1-V_{ij}^{(1)}\}U_{ij}^{(1)}
+V_{ij}^{(1)} W_{ij}\nonumber\\
A_{ij}^{(2)}&=\{1-V_{ij}^{(2)}\}U_{ij}^{(2)}+V_{ij}^{(2)} W_{ij}\nonumber\\
U_{ij}^{(k)}\,|\,z&\overset{d}{=}{\mathrm{Bern}}\{ \theta_{z_i z_j}\},\quad V_{ij}^{(k)}\,|\,z\overset{d}{=}{\mathrm{Bern}}\{ r_{z_i z_j}\},\;k=1,2,\nonumber\\
W_{ij}\,|\,z&\overset{d}{=}{\mathrm{Bern}}\{\theta_{z_i z_j}\},
\end{align}
where $U_{ij}=(U^{(1)}_{ij},U^{(2)}_{ij})^{T}$ and $V_{ij}=(V^{(1)}_{ij},V^{(2)}_{ij})^{T}$ each
contain two (independent) Bernoullis and $W$ is a scalar Bernoulli. The label vector $z=(z_{1},\ldots,z_{n})^{T}$ follows a multinomial distribution (as in the standard one-dimensional SBM).
Note that we can only generate Bernoullis for $V_{ij}^{(k)}$ if the specified mean $r_{ab}$ is non-negative.
Now, clearly, conditionally on $z_i=a$ and  $z_j=b$, the above mechanism leads to
\begin{align*}
\E \{A_{ij}^{(1)}\,|\,z\}&=\{1-r_{ab}\}\theta_{ab}
+r_{ab} \theta_{ab}=\theta_{ab}\\
\E\{A_{ij}^{(2)}\,|\,z\}&=\{1-r_{ab}\}\theta_{ab}+r_{ab} \theta_{ab}=\theta_{ab}\\
\cov\{A_{ij}^{(1)},A_{ij}^{(2)} \,|\,z\}&=
\cov\{\{1-V_{ij}^{(1)}\}U_{ij}^{(1)}
+V_{ij}^{(1)} W_{ij},\{1-V_{ij}^{(2)}\}U_{ij}^{(2)}+V_{ij}^{(2)} W_{ij}\}.
\end{align*}
To understand second moments we note that 
\begin{align*}\cov\{A_{ij}^{(1)},A_{ij}^{(2)} \,|\,z\}&=\var\{W_{ij} \}\E \{ V_{ij}^{(1)}\}\E \{ V_{ij}^{(2)}\}\\
&=\theta_{ab}\{1-\theta_{ab}\}r_{ab}^2.
\end{align*}
This allows us to generate positively correlated random variables, as if we assume that $z_{i}=a$ and $z_{j}=b$, then ${\mathrm{corr}}\{A_{ij}^{(1)},A_{ij}^{(2)}\}=r^2_{ab}.$
Generating negatively correlated Bernoullis is relatively challenging but possible, and requires a different scheme to realize \citep[e.g.][]{andel2010bernoulli}, already adressed in time series.

\subsection{Correlated Homogeneous Graph Limit Model}\label{subsec:corrhomgraphlimit}
The correlated stochastic blockmodel clearly provides a
simplified description for a pair of networks via a block-constant coherence
matrix $R$. In this coherence matrix, the diagonal and off-diagonal blocks
correspond to coherence between graphs, for pairs of node within blocks and
across blocks, respectively. The easiest extension to graph limits is to
directly follow the stochastic blockmodel and replace the blockmodel
specification by the graph limit functions $f^{(1)}(x,y)$ and
$f^{(2)}(x,y)$
and a scale factor $\rho_n$ as in \eqref{eqn:AHthmsp},
coupled with a coherence function {$r_{0}(x,y)$} as defined
by~\eqref{eqn:coherence}.  
As in our specification of the correlated stochastic blockmodel, assume that $f^{(1)}(x,y)=f^{(2)}(x,y)=f(x,y)$, so that marginally $A_{ij}^{(l)}\,| \,\xi\, \sim 
{\mathrm{Bern}}\left\{ \rho_n f(\xi_i,\xi_j) \right\}$, $l=1,2$. Then given a (positive) correlation function $r(x,y)$, the corresponding correlated pair of networks follow via the scheme given in \eqref{eqn:gen}, except with $\xi$ instead of $z$; $\rho_{n}f(\xi_i,\xi_j)$ instead of $\theta_{z_{i}z_{j}}$ and $\rho_n \sqrt{r(\xi_i,\xi_j)}$ instead of $r_{z_{i}z_{j}}.$
This then produces a $2$-layer correlated homogeneous graph limit model with the appropriate graph limits (please refer to calculations of the previous section).

\subsection{Blockmodel to graph limits}
The general $2$-layer correlated stochastic blockmodel may be studied as it is or via a graph limit parameterization. To construct the corresponding graph limit objects we proceed as follows.
Define $H_0=0$, and iteratively the set of constants $H_a=\sum_{k=1}^a h_k, a \in \{1,\ldots,K\}$.
Then for $l\in\{1,2\}$, we define
\begin{align}
f^{(l)}(x,y)&=\sum_{a,b=1}^k\theta^{(l)}_{ab}\I\left(x\in (H_{a-1}/n,H_a/n]\right) \I\left(y\in (H_{b-1}/n,H_b/n]\right),
\label{eqn:fj}
\end{align}
as the (marginal) graph limit associated with graph $l$.
The Hadamard product has expectation $\varrho_{ab}$. To ensure a well-defined coherence quantity, we assume that $0\leq \varrho_{ab}$. This cross-moment leads us to the cross-graph limit, which is given by
\begin{align*}\label{eqn:fj12}
f^{(12)}(x,y)&=\sum_{a,b=1}^k\varrho_{ab}\I\left(x\in (H_{a-1}/n,H_a/n]\right) \I\left(y\in (H_{b-1}/n,H_b/n]\right).
\end{align*}
This corresponds to a bivariate representation of the blockmodel. To go from $\varrho_{ab}$ to the coherence we define
\begin{equation*}
r(x,y)=\sum_{a,b=1}^k\frac{\varrho_{ab}-\theta^{(1)}_{ab}\theta^{(2)}_{ab}}{\sqrt{\theta^{(1)}_{ab}\theta^{(2)}_{ab}(1-\theta^{(1)}_{ab})(1-\theta^{(2)}_{ab})}}\I\left(x\in (H_{a-1}/n,H_a/n]\right) \I\left(y\in (H_{b-1}/n,H_b/n]\right).
\end{equation*}
\subsection{Correlated Inhomogeneous Graph Limit Model}\label{subsec:corrgraphlimit}

Given a pair of graphs $(\mathcal{G}_{1},\mathcal{G}_{2})$, observed on the same set of nodes, a common question of interest is the similarity between them, or more generally, whether the presence of an edge between nodes $i$ and $j$ in $\mathcal{G}_{1}$ implies the corresponding edge in $\mathcal{G}_{2}$. With this in mind, and proceeding in an analogy with what is known in signal processing, as an ``input--output'' relationship, we consider the following approach to modeling relationship between two graphs:
\begin{equation}\label{eqn:rel12}
A_{ij}^{(2)}\,|\, \alpha, \,\xi,\,A_{ij}^{(1)} \sim 
{\mathrm{Bern}}\left\{  h(\xi_i,\xi_j)A_{ij}^{(1)}\right\},\quad 1\leq j<i\leq n,
\end{equation}
where we may interpret $A_{ij}^{(1)}$ as an input to obtain an output  $A_{ij}^{(2)}$. This may also be viewed as ``modulation'' as it is similar to amplitude and frequency modulation in signal processing, being a multiplicative process. 
Note that \eqref{eqn:rel12} implies that 
$A_{ij}^{(2)}=1 \implies A_{ij}^{(1)}=1$, so 
$\mathcal{G}_2 \subseteq \mathcal{G}_1$.

Let $f^{(1)}(x,y)$ and $f^{(2)}(x,y)$ denote graph limit functions corresponding to $A^{(1)}$ and $A^{(2)}$ (e.g. \eqref{eqn:AHthmsp}) and let $r(x,y)$ denote the limiting correlation function. Let $\rho_n f^{(1)}(\xi_i,\xi_j)=p_{ij}^{(1)}.$ 
To parameterize $A_{ij}^{(2)}$ given $A_{ij}^{(1)}$, we calculate its expectation using the law of iterated expectation, which is given by
\begin{align*}
\E\{ A_{ij}^{(2)}\,|\, \,\xi\}
&=\E_{A_{ij}^{(1)}\,|\,  \,\xi}\{
\E\{ A_{ij}^{(2)}\,|\, \,\xi,\,A_{ij}^{(1)}\}\}=\E_{A_{ij}^{(1)}\,|\,  \,\xi}\{
h(\xi_i,\xi_j)A_{ij}^{(1)}\}\\
&=\E_{A_{ij}^{(1)}\,|\,\,\xi}\{
\rho_n h(\xi_i,\xi_j) f^{(1)}(\xi_i,\xi_j)\}=
\rho_n h(\xi_i,\xi_j) f^{(1)}(\xi_i,\xi_j)\\
&= h(\xi_i,\xi_j) p_{ij}^{(1)}= p_{ij}^{(2)},
\end{align*}
{where we defined $p_{ij}^{(2)}$ as a pointwise product of $h$ and $p_{ij}^{(1)}$}.  Furthermore we find that the expected value of the Hadamard product under \eqref{eqn:rel12} is given by
\begin{align}
\nonumber
\E\biggl\{ \left(A^{(1)}\circ A^{(2)}\right)_{ij}\,|\, \,\xi \biggr\}
&=\E\{ h(\xi_i,\xi_j)A_{ij}^{(1)} \,|\, \,\xi\}=\rho_n
h(\xi_i,\xi_j) f^{(1)}(\xi_i,\xi_j)\equiv p_{ij}^{(12)},\\
&=\Pr\{ A^{(1)}_{ij}=1\cap A^{(2)}_{ij}=1\}.
\end{align}
Clearly, $p_{ij}^{(12)}=p_{ij}^{(2)}$, and thus, the probability of an edge in $\mathcal{G}_{2}$ is equal to the probability of the corresponding edge occuring in both $\mathcal{G}_{1}$ and $\mathcal{G}_{2}$. Thus, $A^{(2)}_{ij}=1$ with a high probability 
 implies a high probability of the edge occuring in both the layers or $A^{(1)}_{ij}=1$.
Also, clearly $p_{ij}^{(1)}=0$, implies $p_{ij}^{(2)}=0$.
 Further implications can be understood by noting the the probabilities associated with the four possible outcomes, as follows
\begin{align*}
\Pr\{ A^{(1)}_{ij}=0\cap A^{(2)}_{ij}=1\}&=\Pr\{  A^{(2)}_{ij}=1\}-\Pr\{ A^{(1)}_{ij}=1\cap A^{(2)}_{ij}=1\}\\
&=\rho_n h(\xi_i,\xi_j)f^{(1)}(\xi_i,\xi_j)-\rho_n
h(\xi_i,\xi_j) f^{(1)}(\xi_i,\xi_j)\\
&=0\\
\Pr\{ A^{(1)}_{ij}=1\cap A^{(2)}_{ij}=0\}&=\Pr\{  A^{(1)}_{ij}=1\}-\Pr\{ A^{(1)}_{ij}=1\cap A^{(2)}_{ij}=1\}\\
&=\rho_n f^{(1)}(\xi_i,\xi_j)-\rho_n
h(\xi_i,\xi_j) f^{(1)}(\xi_i,\xi_j)\\
&=\rho_n f^{(1)}(\xi_i,\xi_j)\{1-h(\xi_i,\xi_j)\}\\
\Pr\{ A^{(1)}_{ij}=0\cap A^{(2)}_{ij}=0\}&=1-\rho_n
h(\xi_i,\xi_j) f^{(1)}(\xi_i,\xi_j)-\rho_n f^{(1)}(\xi_i,\xi_j)\{1-h(\xi_i,\xi_j)\}\\
&=1-\rho_n f^{(1)}(\xi_i,\xi_j).
\end{align*}
Note that the scaling in this model is not homogeneous as the probability of an edge in the second graph will scale like the probability of having the corresponding edge in both the graphs, which the generating mechanism in a sense is regulating.
The conditional correlation under this model, takes the form
\begin{align}
\nonumber
r(\xi_{i},\xi_{j})&=\frac{p_{ij}^{(12)}- p_{ij}^{(1)} p_{ij}^{(2)}}{[\var\{ A_{ij}^{(1)}\}\var\{ A_{ij}^{(2)}\}]^{1/2}},\quad 1\leq j<i\leq n,
\\
& 
=\left[\frac{
h(\xi_i,\xi_j)\{1 -\rho_n f^{(1)}(\xi_i,\xi_j)\}}
{ 1-\rho_n h(\xi_i,\xi_j) f^{(1)}(\xi_i,\xi_j) }\right]^{1/2}
.
\end{align}
Clearly, as $\rho_{n}\rightarrow 0$, this is a finite order one
  quantity with the leading term given by
  $h(\xi_i,\xi_j)^{1/2}$. Thus, coherence $r_{0}(\xi_{i},\xi_{j})$ under this model is set equal to the conditional correlation $r(\xi_{i},\xi_{j})$.
This is in contrast to
  what we observed under the general model described in
  \cref{sec:scaledgraphs} and shows that
 the choice of scaling has important implications. Naturally still the Cauchy--Schwartz inequality applies in this setting and so the magnitude of $r(\xi_{i},\xi_{j})$ is bounded.

\section{Discussion}\label{sec:disc}
Introducing correlation between graphs is a natural next step in modelling graphs, thereby acknowledging dependence also in this non-Euclidean setting. {This clearly goes beyond the mean of a graph, a topic with significant interest~\citep{lunagomez2020modeling}.
Graphs can naturally also be dependent without being perfectly replicated. In graph theory the notion of graph limits were partially introduced to use tools from analysis in a combinatorics setting.
{Our framework shows how simple `linear' operations may be interpreted in the space of graph limits.}
Further, we note that marginally the edge variables in a single  graph are correlated when one of the indices are fixed to be the same. Edges in exchangeable models, and scaled exchangeable arrays, can only exhibit positive correlation inside a single exchangeable graph~\citep[p.133]{aldous}. Once we move between two ``layers'' or graphs on the same nodes, negative correlation can be generated. One simple way of doing so would be to take $f^{(1)}(x,y)=g(x,y)$ and $f^{(2)}(x,y)=1-c \cdot g(x,y)$ for some positive $c$ and function $g(x,y)$, \citep[e.g.][]{chaganty2006range}.
Of course as the dimension of the Bernoulli vector increases data structures become more difficult to characterise~\citep{huber2017bernoulli}. {Given this, for multiplex networks with more than two layers, one may proceed by applying the proposed framework to all possible pairs of network layers. Thus, reducing a $d$-dimensional model to $d(d-1)/2$ two-dimensional models parameterized by edge coherence. Edge coherence between a pair of networks is a weighted graph itself. It summarizes their dependence structure and may be used for key applications such as detecting changes in second-order dependence structure across network layers observed over time or space.}

\section{Acknowledgements}\label{sec:ack}
 This work was supported by the European Research Council under Grant CoG 2015-682172NETS,
 within the Seventh European Union Framework Program 
 and the Knut and Alice Wallenberg Foundation.
\bibliographystyle{apalike}
\bibliography{coherence}
\end{document}